\documentclass[preprint,aps]{revtex4}

\usepackage{graphicx}
\usepackage{dcolumn}
\usepackage{bm}

\begin{document}


\title{
Group Chase and Escape with Conversion from Targets to Chasers 
}

\author{Ryosuke Nishi$^{1,2}$}
 \email{tt097086@mail.ecc.u-tokyo.ac.jp}
\author{Atsushi Kamimura$^{3}$}
\author{Katsuhiro Nishinari$^{4}$}
\author{Toru Ohira$^{5}$}

\affiliation{
$^{1}$ Department of Aeronautics and Astronautics,
School of Engineering, The University of Tokyo,
Hongo, Bunkyo-ku, Tokyo 113-8656, Japan.
\\
$^{2}$ Japan Society for the Promotion of Science, 
Kojimachi, Chiyoda-ku, Tokyo 102-8471, Japan.
\\
$^{3}$ Institute of Industrial Science, The University of Tokyo, 
4-6-1, Komaba, Meguro-ku, Tokyo, 153-8505, Japan.
\\
$^{4}$ Research Center for Advanced Science and Technology, The University of Tokyo,
Komaba, Meguro-ku, Tokyo 153-8904, Japan.
\\
$^{5}$ Sony Computer Science Laboratories, Inc., 
3-14-13, Higashi-gotanda, Shinagawa-ku, Tokyo, 141-0022, Japan.
}

\date{\today}
\begin{abstract}
We are studying the effect of converting caught targets into new chasers in the context of the recently proposed `group chase and escape' problem. Numerical simulations have shown that this conversion can substantially reduce the lifetimes of the targets when a large number of them are initially present. At the same time, it also leads to a non-monotonic dependence on the initial number of targets, resulting in the existence of a maximum lifetime. As a counter effect for this conversion, we further introduce self-multiplying abilities to the targets. We found that the longest lifetime exists when suitable combination of these two effects is created.
\end{abstract}
\maketitle

\section{Introduction}
The collective motions of interacting entities have recently attracted a lot of attention. They include groups such as molecular motors, ants, fish, birds, pedestrians, and vehicles. They are generally termed self-driven particles (SDPs). The universal phenomena in these systems, such as the phase transitions and metastable states, have been discovered and investigated [1-3].

`Chase and Escape', often observed in nature, has long been studied in mathematical literature. Even the simplest of cases, a two particle system composed of one chaser and one target, posed the challenging mathematical problem of analytically describing their trajectories [4-6]. This chase and escape problem can be applied to SDPs and also systems that show chemotaxis [7]. Systems with a number of chasers and one target have been modeled and analyzed [8,9]. Systems with several chasers and targets were investigated in the fields of game theory, mechanical engineering, and multi-agent problems [10,11]. 
This topic has recently been further extended to take into consideration cases where large numbers of chasers and targets exist, which is called ``group chase and escape" [12]. In this model, the chasers pursue the nearest targets, while the targets try to escape from be caught by the nearest chasers to avoid removal. 
%
We should note that in a study of collective motion, pursuit-and-escape interactions were 
also investigated by using Brownian particles not grouped separately as chasers and escapees [13]. 
In contrast, in group chase and escape, particles are grouped separately. 
Some of the properties associated with such group chase and escape were found. For example, spontaneous self-organizations are formed for both the chasers and targets without any internal communications within each group. In addition, given the initial number of targets, the optimal number of chasers can be identified by introducing the running cost. We have extended this model of group chase and escape in this paper by posing to ourselves the following question, ``How do the dynamics change if the caught targets are converted to chasers, instead of removing them?" This conversion can be related to the spread of rabies [14] by interpreting that the chasers are infected animals and the targets are susceptible ones. The targets caught by the chasers become infected, i.e., they are converted to chasers. To answer the above question, we investigated how the lifetimes of targets change as we vary the rate of conversion. This article is composed as follows. Our model with the given conversions is introduced in section \ref{sec:model}. Then, we discuss the numerical simulations and their results in section \ref{sec:simu}. Section \ref{sec:conc} contains our conclusive discussions.
\section{Model}
\label{sec:model}
We took into consideration two kinds of particles: chasers and targets. 
$N_C^0$ chasers and $N_T^0$ targets are initially randomly placed on a $L_x \times L_y$ discrete square lattice with periodic boundaries. We define the discrete time update for the chasers and targets. At each time step, the chasers and targets move according to the following rules. Each chaser at $(x_C,y_C)$ tries to catch its nearest target. The nearest target is a target placed at a minimum distance from the chaser, where the distance between two points $(x_1,y_1)$ and $(x_2,y_2)$ is defined by using the following equation,\\
$\sqrt{\left\{\min[x_1-x_2, L_x-(x_1-x_2)]\right\}^2 + \left\{\min[y_1-y_2, L_y-(y_1-y_2)]\right\}^2}$. \\
If there are two or more targets with the same minimum distance from a chaser, one of them is randomly chosen as its nearest target. The position of the chosen target is denoted by $(x_{Cn}, y_{Cn})$.

\begin{figure}[htbp]
\includegraphics[width=1.0\linewidth]{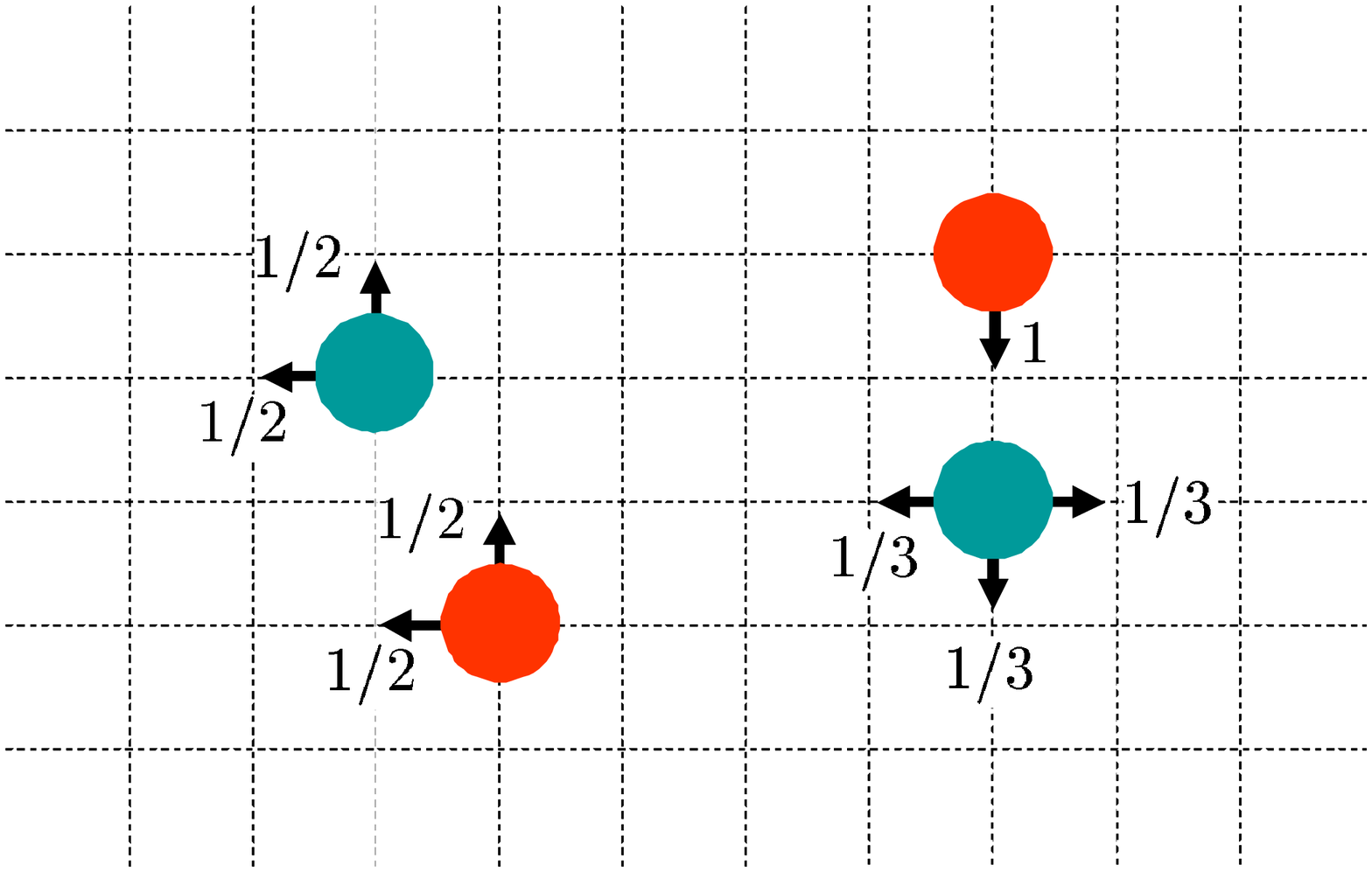}
\caption{Directions of hopping for chasers (red) and targets (green). }
\label{fig:directions}
\end{figure}

\begin{figure}[htbp]
\includegraphics[width=1.0\linewidth]{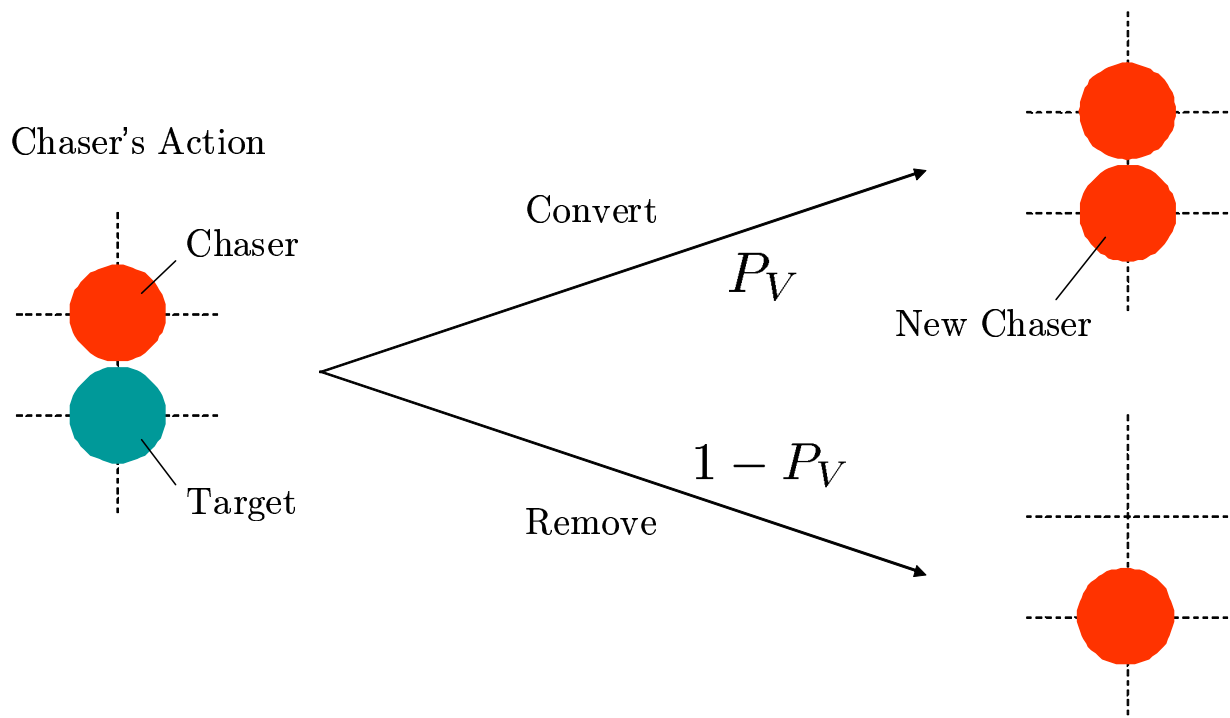}
\caption{A caught target (green) is converted to a new chaser (red) with probability $P_V$.}
\label{fig:conversion}
\end{figure}

The behavior of each chaser changes whether the chosen target is located next to it or not (i.e., the distance is one or larger). If the distance is larger than one, it hops one of its four neighboring cells to decrease the distance, as shown in Figure \ref{fig:directions}. If there is more than one such cell, one of them is randomly chosen using equal probabilities. However, we also assume exclusive volume effects. So, the chaser can hop only if the cell is not occupied by another chaser. If the cell is occupied, it stays at its current position. On the other hand, if the target is located next to the chaser, it is caught. In the previous study [12], this catch event resulted in the removal of the target. In this paper, we introduce the new additional possibility of converting the target into a new chaser, as shown in Fig. \ref{fig:conversion}. `Remove' or `Convert' actions take place with the probabilities $1 - P_V$ and $P_V$, respectively. If a remove action is chosen, the chaser hops to the position of its nearest target and the target disappears. If a convert action is chosen, the chaser does not move and the nearest target is converted into a new chaser. We note that $P_V=0$ corresponds to a situation where all the caught targets are removed leading to the original model [12]. $P_V=1$ represents the case where all the caught targets become chasers. The movements of the targets are as follows. A target at $(x_T,y_T)$ tries to hop one cell to escape from its nearest chaser at $(x_{Tn},y_{Tn})$, and if there are multiple chasers with the minimum distance to the target, one chaser is randomly chosen as the nearest chaser. Similar to the chasers' rule, the target hops one of its four neighboring cells to increase the distance, as shown in Fig. \ref{fig:directions}. If there is more than one such cell, one of them is randomly chosen using equal probabilities. 
We also assume exclusive volume effects for the targets. 
We use the following overall discrete-time updating schedule.  
In each time step, all the chasers are moved in a random order first, 
and then, all the targets are updated using a random order. 
\section{Simulations}
\label{sec:simu}
We investigated the entire lifetime $\cal T$ and typical lifetime $\tau$ for various conversion probabilities, $P_V$, using numerical simulations to evaluate the effect of the conversion rule. $\cal T$ is defined as the period from the start time $t=0$ to the time when the last target is caught. $\tau$ is defined as $\sum_{i=1}^{\cal T} i (N_T^{i-1} - N_T^i) / N_T^0$.

\begin{figure}[htbp]
\includegraphics[width=1.0\linewidth]
{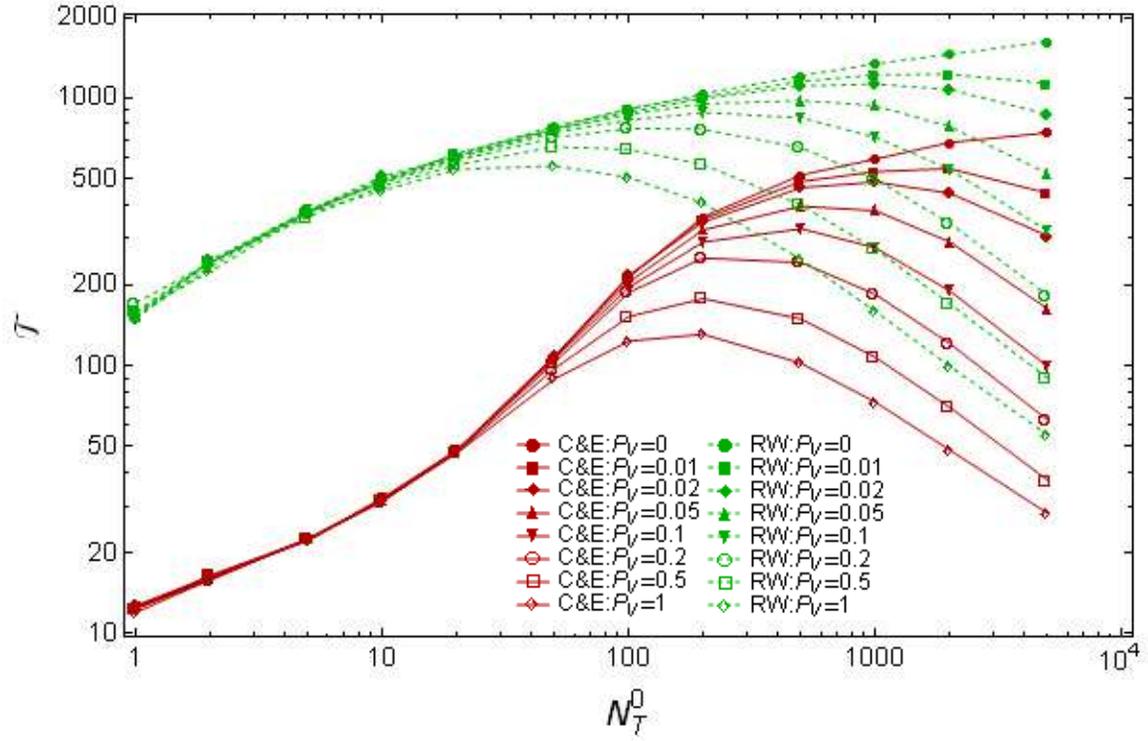}
\caption{Relationships between $\cal T$ and $N_{T}^{0}$ with various $P_V$ for the two cases, 
the chasing and escaping case (C\&E) and the case of random walks of chasers and targets (RW). 
}
\label{fig:T_vs_Nt0}
\end{figure}

\begin{figure}[htbp]
\includegraphics[width=1.0\linewidth]
{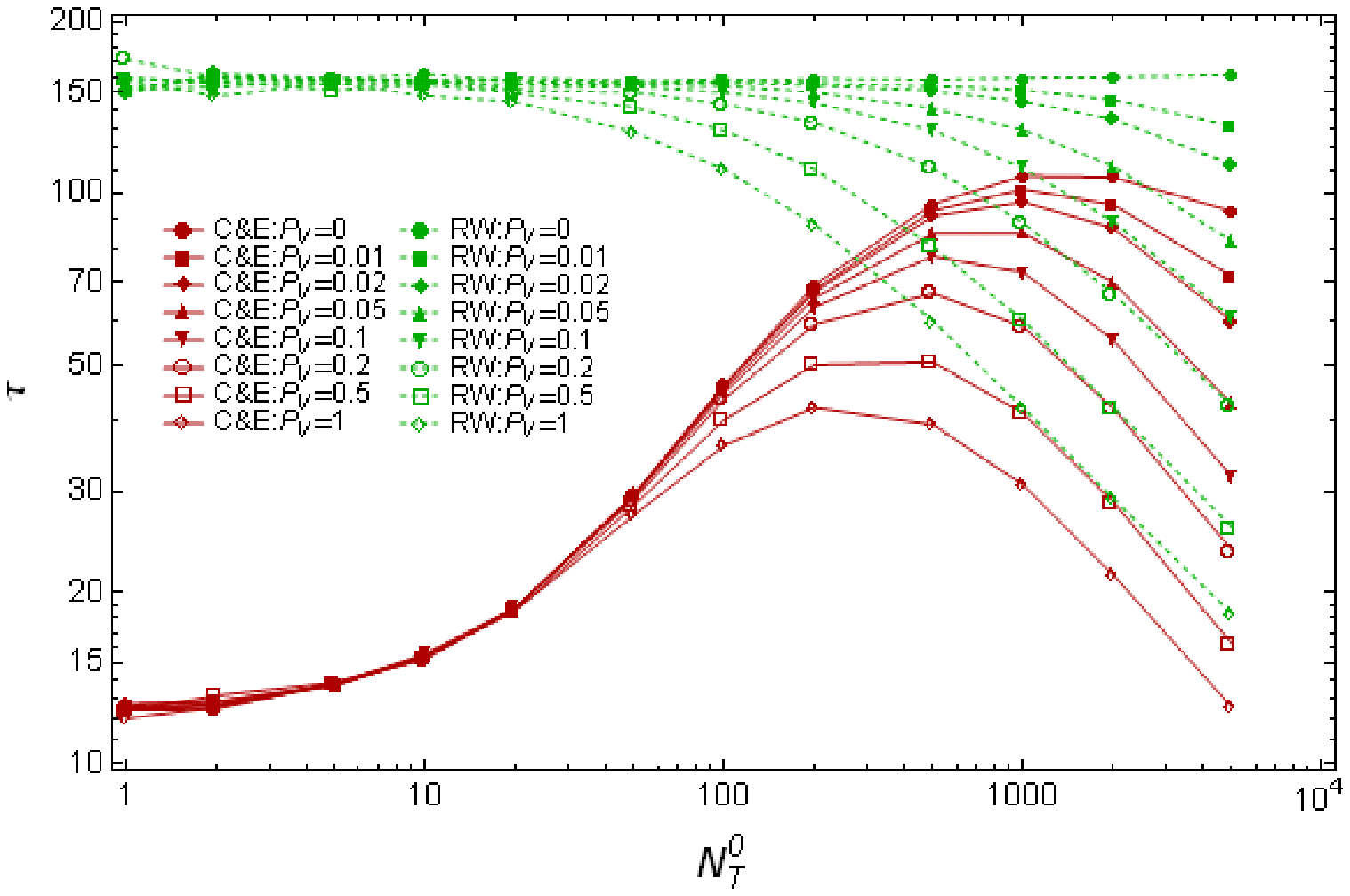}
\caption{Relationships between $\tau$ and $N_{T}^{0}$ with various $P_V$.}
\label{fig:tau_vs_Nt0}
\end{figure}

We plotted $\cal T$ and $\tau$ versus $N_T^0$ 
with $P_V \in \left\{0, 0.01, 0.02, 0.05, 0.1, 0.2, 0.5, 1 \right\}$,
in Figs. \ref{fig:T_vs_Nt0} and \ref{fig:tau_vs_Nt0}, respectively. The following conditions were given. The size of the system is $L_x = L_y = 100$, the initial number of chasers is $N_C^0=100$, and the results are averaged over $1000$ samples. 
In addition to the chasing and escaping rule defined in section \ref{sec:model}, 
we also present the results for a case in which all the chasers and targets 
obey random walks for a comparison. 

Maxima $\cal T$ exist for the plots in the $P_V \ge 0.01$ range 
for both the chasing and escaping rule  
and the random walking rule. However, there is no observed maximum $\cal T$ for $P_V=0$. As for $\tau$, the maxima $\tau$ exists for all the $P_V$ 
for the chasing and escaping rule. 
For the random walking rule, however, $\tau$ stays almost constant at $P_V=0$ and monotonically decreases with $N_T^0$ 
for $Pv \ge 0.01$. 

The existence of a maximum $\cal T$ can be explained as follows. In the original group chase and escape model ($P_V = 0$) [12], $\cal T$ increases as $N_T^0$. If $N_T^0$ is small, the conversion $P_V > 0$ would not be relevant. However, as $N_T^0$ increases, the chasers frequently catch targets and convert them to new chasers, which results in the fast spread of a chain of conversions. This spread would drastically decrease the entire lifetime $\cal T$. Thus, these two factors lead to the maximum $\cal T$ as a function of $N_T^0$. This would also be applied to the existence of a maximum $\tau$ 
for the chasing and escaping rule.  
The latter factor also causes the monotonic decrease $\tau$ for the random walking rule. Note that a maximum $\tau$ in $P_V=0$ was reported in [12]. Furthermore, it is noteworthy that $\cal T$ with $N_T^0=1$ is larger than $\cal T$ with $N_T^0=5000$ for a large $P_V$ for the random walking case. 

We also plot $\cal T$ and $\tau$ as a function of $N_C^0$. The initial number of targets is fixed at $N_T^0=100$, the degree of conversion is given as $P_V \in \left\{0, 0.2, 0.5, 1 \right\}$ and the other conditions are the same in Figs. \ref{fig:T_vs_Nt0} and \ref{fig:tau_vs_Nt0}. 

\begin{figure}[htbp]
\includegraphics[width=1.0\linewidth]
{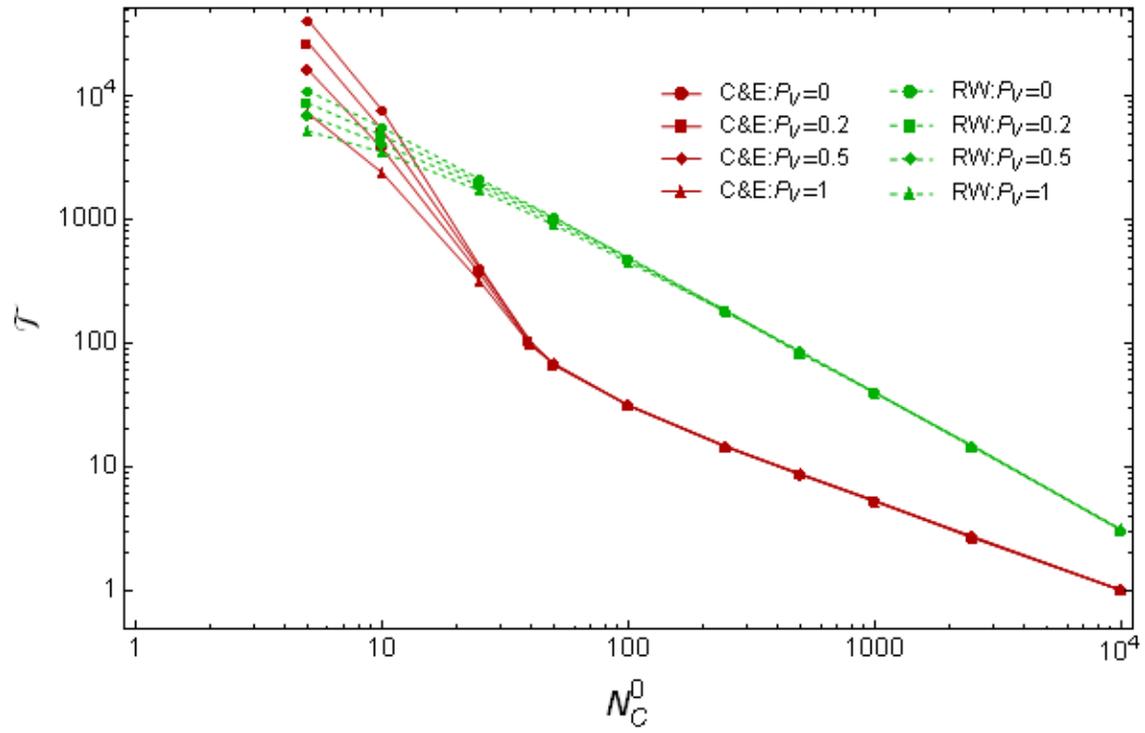}
\caption{Relationships between $\cal T$ and $N_{C}^{0}$ with various $P_V$. 
The notations C\&E and RW correspond to the chasing and escaping case 
and the case of random walks of chasers and targets, respectively.}
\label{fig:T_vs_Nc0}
\end{figure}

\begin{figure}[htbp]
\includegraphics[width=1.0\linewidth]
{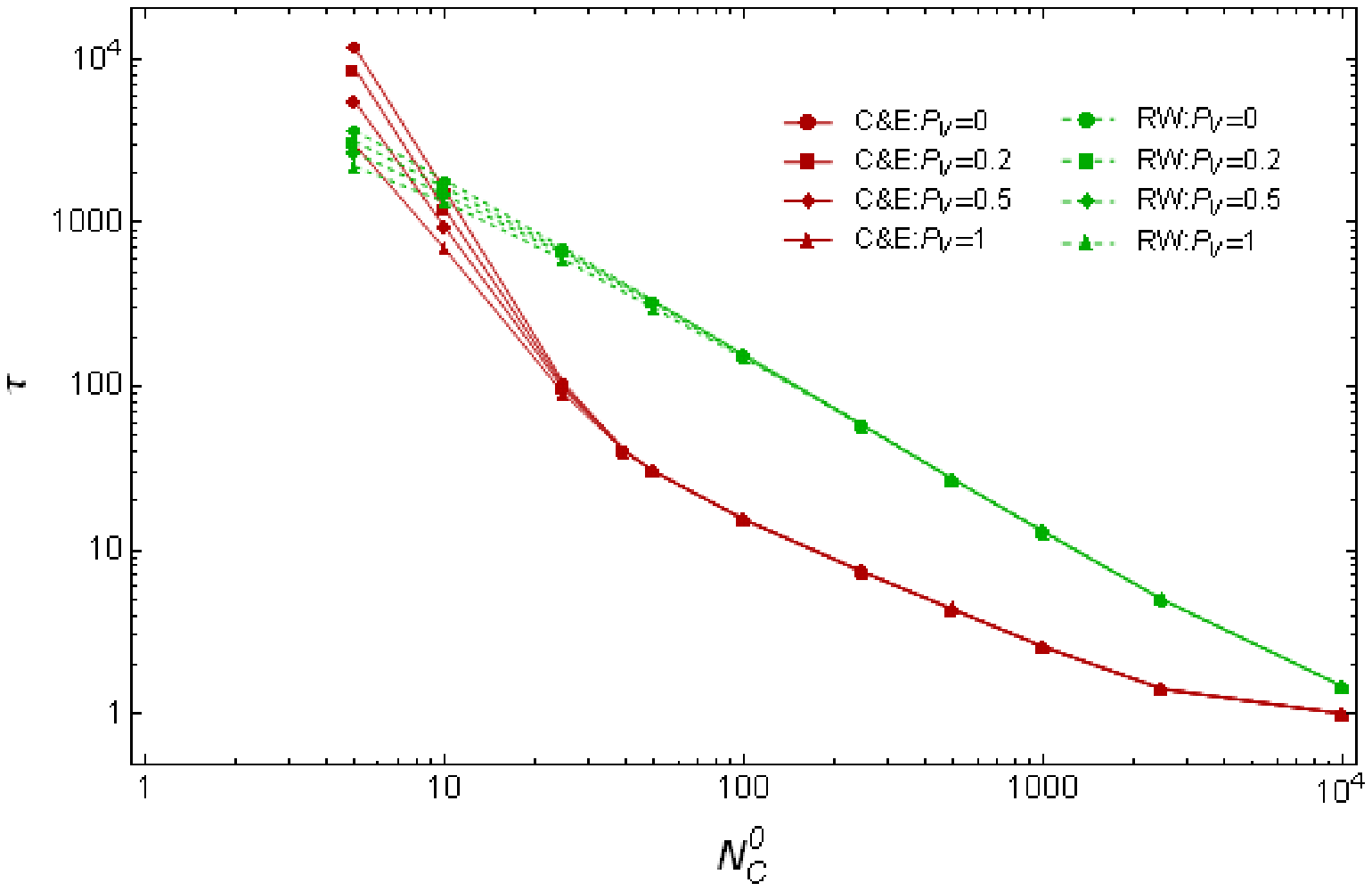}
\caption{Relationships between $\tau$ and $N_{C}^{0}$ with various $P_V$.}
\label{fig:tau_vs_Nc0}
\end{figure}

The $\cal T$ and $\tau$ results versus $N_C^0$ are shown in Figs. \ref{fig:T_vs_Nc0} and \ref{fig:tau_vs_Nc0}, respectively. For the chasing and escaping cases, 
a $\cal T$ with various 
$P_V$ merges around $N_C^0=40$ and a $\tau$ also merges around $N_C^0=40$. 
The lines do not depend on $P_V$ above the critical points. Meanwhile, there is a spread in the slope below the points. Still, in the case of chase and escape, we observed clear changes in the slopes around the points, while in the random walk case they smoothly change. 

Such clear changes 
for the chasing and escaping rule 
were reported for $P_V=0$ [12]. This difference reflects the sudden changes in effectiveness of catching the targets. That means there is a critical number of chasers, beyond which adding more chasers is not as effective. Our results indicate that this critical number also exists even with a conversion, $P_V > 0$. In addition, this understanding is consistent with the result showing that the critical point does not depend on $P_V$, but is dictated by a large number of chasers. 

In the above simulations, the targets monotonically decrease due to the catch-up and are bound to extinction. In addition, the conversions from targets to chasers are more advantageous for the chasers. Here, we consider giving the targets the ability to proliferate in order to resist against extinction. The details for this new rule are given as follows. When each target hops to one of its neighboring sites in a single time step, a new target arises on the original site with the probability $P_T$. $P_T$ is a parameter for the natural increase. $P_T=0$ is the case in which the targets do not increase, while $P_T=1$ is the case in which the targets always increase when they move. Note that this increasing rule takes into account the exclusive volume effect. When the site is filled with a target, a new target does not arise due to the shortage of space. Note also that the new target stays in the time step. 

\begin{figure}[htbp]
\includegraphics[width=1.0\linewidth]
{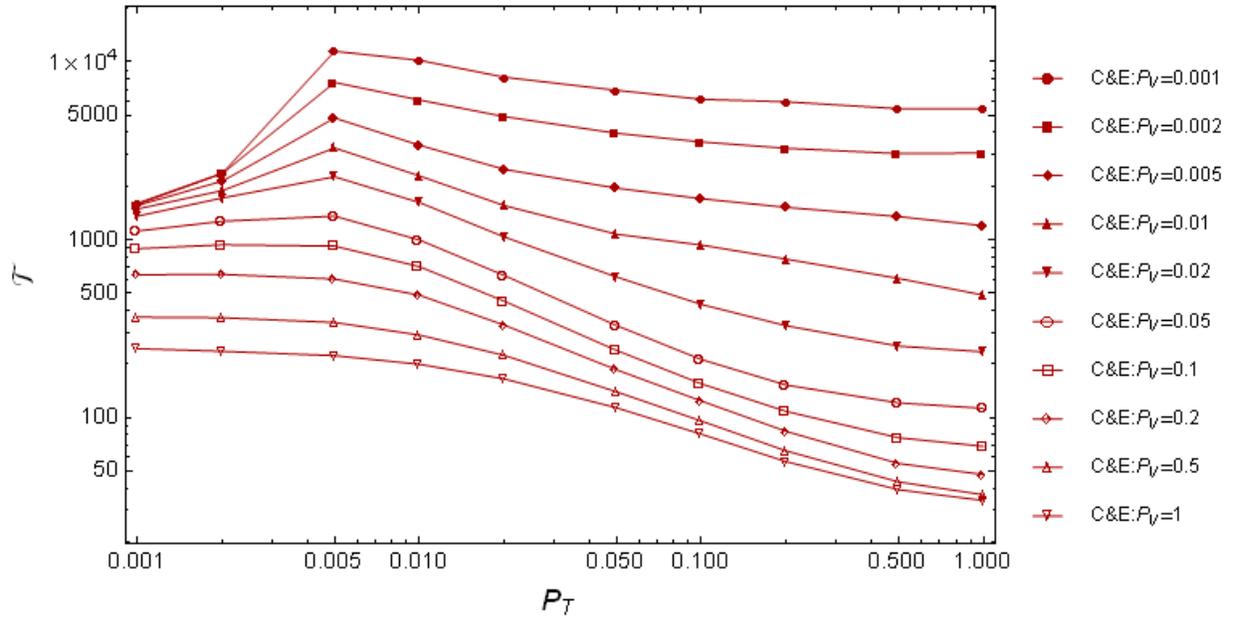}
\caption{Relationships between $\cal T$ and $P_T$ with various $P_V$ 
in the chasing and escaping case.}
\label{fig:CCTE_T_vs_Pt_Pc=0}
\end{figure}

\begin{figure}[htbp]
\includegraphics[width=1.0\linewidth]
{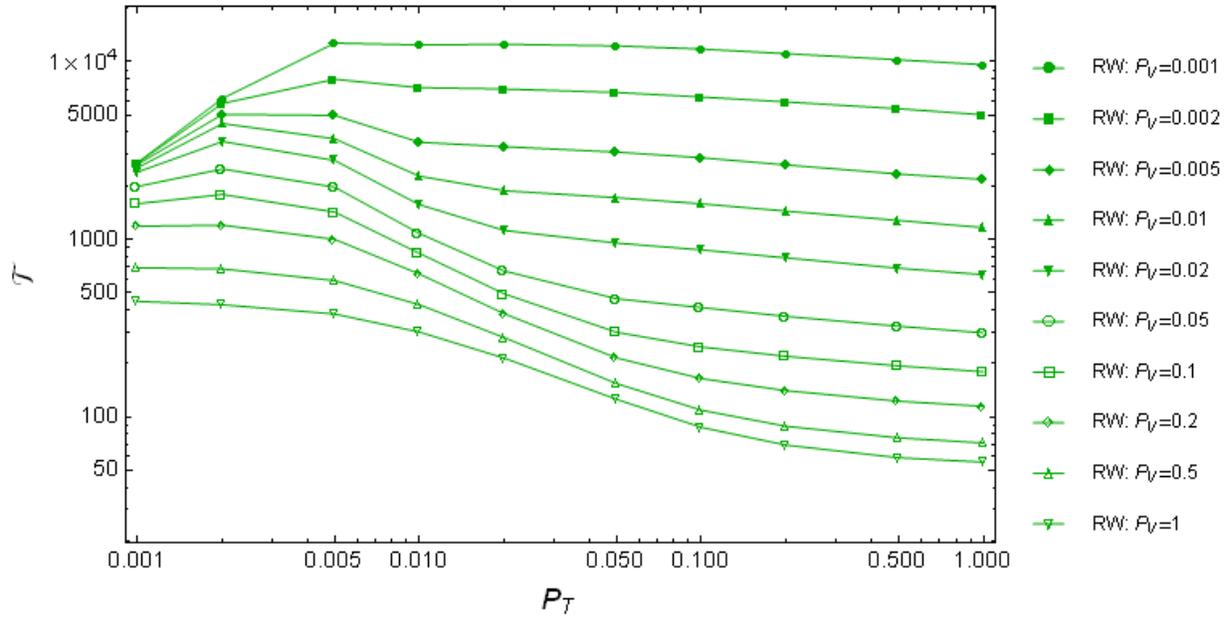}
\caption{Relationships between $\cal T$ and $P_T$ with various $P_V$ 
in the case of random walks of chasers and targets.}
\label{fig:CRWTRW_T_vs_Pt_Pc=0}
\end{figure}

We plot the relationships between the lifetime $\cal T$ and $P_T$ for various $P_V$. The conditions for this are given as follows. The size of the system is $L_x=L_y=50$, the initial numbers are $N_C^0=10$ and $N_T^0=50$, 
and the simulations are performed for $1000$ samples. 
The results are shown in Figs. \ref{fig:CCTE_T_vs_Pt_Pc=0} and \ref{fig:CRWTRW_T_vs_Pt_Pc=0} for the chasing and escaping and random walking rules, respectively. In both figures, there is a maximum of $\cal T$ for a small $P_V$. 

The mechanism for producing such a maximum $\cal T$ can be explained as follows. If $P_T$ is very small, all the targets are caught by the chasers before they can proliferate. As $P_T$ increases, the targets have more chance to increase, which leads to an increase in $\cal T$. With $P_T$ increasing, however, the converted chasers make $\cal T$ smaller. In this case, the targets initially explosively increase, and afterward a number of targets are caught and converted to new chasers. Due to this increase in converted chasers, the catching rates increase, and finally the number of targets start to very quickly decrease. The existence of a maximum $\cal T$ for a small $P_V$ suggests that the proliferation rate of the targets should not be too high for longer survival. On the other hand, if $P_V$ is large, the conversion ability of the chasers overrides the targets' proliferation abilities, which obscures the maximum. 
\section{Conclusive Discussions}
\label{sec:conc}
We proposed group chasing and escaping models with a conversion rule from targets to chasers. We found, through numerical simulations, that the conversion rule provide the optimum initial number of targets to produce the maximum lifetime $\cal T$ for the targets. While this maximum is also observed in the random walk rule, the conversion does not affect the qualitative difference with respect to the changes in the initial number of chasers. We also performed simulations using the effect of self-proliferation of the targets. The simulation results again exhibited the maximum lifetime at the appropriate level of proliferation when the conversion rate was moderate. 

We will take the followings into consideration as future works. 
First, we need to explore whether real data exhibit the qualitative behavior of the model. 
Second, we want to take into consideration a time delay element, which in reality may be relevant in infective processes, such as a conversion delay. Third, we would also like to extend our model to include a variety of target resistances, which in turn would be beneficial for developing more efficient strategies for the chasers.

\section{Acknowledgement}
\label{sec:ack}
We would like to thank Dr. Shigenori Matsumoto for his critical advise and helpful discussions. 

\newpage
\newpage
\newpage
\newpage
\newpage
\newpage
\newpage
\newpage

\newpage

\end{document}